\documentclass[%
 reprint,
 amsmath,amssymb,
 aps,
]{revtex4-2}

\usepackage{graphicx}
\usepackage{dcolumn}
\usepackage{bm}

\begin{document}

\preprint{APS/123-QED}

\title{Optimizing Polarizability Distributions for Metasurface Apertures with Lorentzian-Constrained Radiators}

\author{Patrick T. Bowen, Michael Boyarsky, Laura M. Pulido-Mancera}
\author{David R. Smith}%
 \email{drsmith@duke.edu}
\affiliation{%
 Center for Metamaterials and Integrated Plasmonics, Department of Electrical and Computer Engineering,\\ Duke University, Durham, NC 27708, USA
}%

\author{Okan Yurduseven}
\affiliation{
 Centre for Wireless Innovation, Institute of Electronics, Communications and Information Technology, School of Electronics, Electrical Engineering and Computer Science,\\ Queen’s University Belfast, Belfast BT3 9DT, United Kingdom
}%
\author{Mohsen Sazegar}
\affiliation{%
 Kymeta Corporation, 12277 134th Court Northeast, Redmond, Washington 98052, USA
}%

\date{\today}

\begin{abstract}
We present a design strategy for selecting the effective polarizability distribution for a metasurface aperture needed to form a desired radiation pattern. A metasurface aperture consists of an array of subwavelength metamaterial elements, each of which can be conceptualized as a radiating, polarizable dipole. An ideal polarizability distribution can be determined by using a holographic approach to first obtain the necessary aperture fields, which can then be converted to a polarizability distribution using equivalence principles. To achieve this ideal distribution, the polarizability of each element would need to have unconstrained magnitude and phase; however, for a single, passive, metamaterial resonator the magnitude and phase of the effective polarizability are inextricably linked through the properties of the Lorentzian resonance, with the range of phase values restricted to a span of at most 180 degrees. Here, we introduce a family of mappings from the ideal to the available polarizability distributions, easily visualized by plotting both polarizabilities in the complex plane. Using one of these mappings it is possible to achieve highly optimized beam patterns from a metasurface antenna, despite the inherent resonator limitations. We introduce the mapping technique and provide several specific examples, with numerical simulations used to confirm the design approach. 

\end{abstract}

\maketitle


\section{Introduction}

Waveguide-fed metasurface antennas and apertures have steadily gained traction in recent years for a variety of applications, including computational imaging \cite{hunt2013science,hunt2014metaimager,8718481,lipworth2013JOSA, Marks:17,7452548, doi:10.1063/1.4935941,Fromenteze:17,8023980}, communications \cite{Johnson2014,johnson2015sidelobe,grbic2002leakywave,stevenson2016mtenna}, radar and synthetic aperture radar imaging \cite{boyarsky2017sar,pulido2016rma}, and wireless power transfer\cite{doi:10.1063/1.4973345,PhysRevApplied.11.054046,8280564}. A subset of the broader class of metasurfaces \cite{htchen, holloway_review}, waveguide-fed metasurfaces consist of arrays of metamaterial elements distributed over a surface, each considerably smaller than the operating wavelengths and fed by a guided wave. Each subwavelength metamaterial element scatters the incident field predominantly as a polarizable electric or magnetic dipole \cite{sleasman2015element,sleasman2016DyAp2}, introducing both a phase shift and an attenuation to the exciting field. The scattered phase shift and attenuation imparted to the incident wave relate to the geometry of the metamaterial element and are not independent; rather, they are related by the inherent properties of the Lorentzian resonance. Despite the inherent limitations of individual metamaterial resonators, a judicious design strategy for the available polarizability values can provide excellent performance from the simplified metasurface architecture \cite{Landy,PhysRevApplied.8.054048}. Our aim here is to provide a family of mappings from the ideal, unconstrained polarizabilities---needed to produce an arbitrary radiation pattern---to the available, Lorentzian-constrained polarizabilities that optimize the wavefront shaping capabilities. 

In a metasurface antenna either excited by an obliquely incident plane wave or fed by a guided wave, the lack of independent control over phase and amplitude can be largely compensated for by the phase accumulation of the incident wave---similar to the mechanism in leaky-wave or traveling-wave antennas \cite{leaky_wave}, as well as more recently investigated composite right-handed left-handed (CRLH) antennas \cite{crlh}. When the spacing between metamaterial elements is significantly subwavelength ($~\lambda/10$), the incident wave can be sampled using simple "on/off" elements that either transmit or block radiation. For such extreme, subwavelength sampling, high fidelity beams and other radiation patterns can be produced using numerical optimization techniques combined with a semi-analytic forward model \cite{mesa}. When the phase accumulation of the incident wave is coupled with the additional phase shift imparted by the metamaterial elements, a holographic design approach can be implemented. In this scenario, the extreme subwavelength spacing of the metamaterial elements can be relaxed to some extent given the additional freedom associated with the resonance.

\begin{figure}[h]
\centering
\includegraphics[width=8cm]{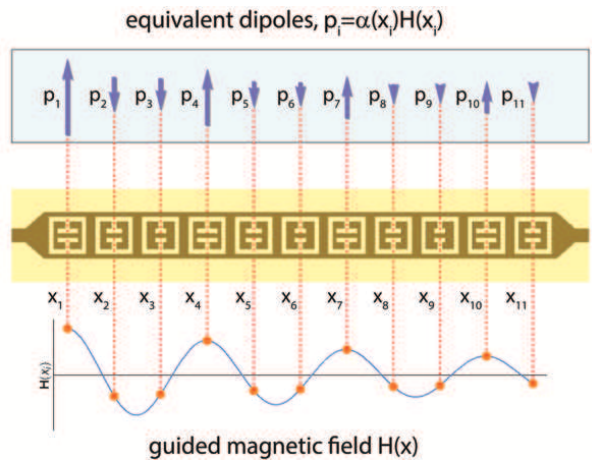}
\caption{Illustration of the waveguide-fed metasurface antenna. A guided wave feeds the metamaterial elements patterned in the upper conductor. Each metamaterial element scatters to good approximation as a polarizable dipole, such that the radiation can be modeled as the sum from a discrete set of dipoles.}\label{fig:dma_architecture}
\end{figure}

One particular implementation of a waveguide-fed metasurface antenna consists of a an array of complementary resonant metamaterial elements patterned into one of the conducting walls of the waveguide (Fig. \ref{fig:dma_architecture}). This waveguide-fed metasurface antenna architecture was analyzed in \cite{PhysRevApplied.8.054048}, where its basic operation and characteristic features were introduced. The confined, guided wave serves as the reference wave, the radiated field constitutes the object wave, and the complementary metamaterial elements comprise the hologram. Since the waveguide mode also introduces a phase shift between subsequent metamaterial elements due to propagation, the waveguide-fed metasurface antenna can reject unwanted diffracted orders that often accompany holograms. For example, there is no radiating zeroth-order beam, as would occur under excitation by an incident plane wave. In \cite{PhysRevApplied.8.054048} it was further shown that an effective array factor could be defined for the aperture, similar to that used for phased arrays, from which directivity patterns could be calculated. It was further shown that the analytic form of the array factor suggested a possible mapping between the ideal and constrained polarizability values, which was used as a design approach to demonstrate beam forming. The mapping found in \cite{PhysRevApplied.8.054048} is found here to be a special case of a family of similar linear mappings in the complex plane.

Compared with active devices, such as phase shifters and amplifiers, the passive, single-resonator metamaterial elements lead to a simpler architecture but with a significantly more constrained design space \cite{hansen2009phasedarray}. Tuning the resonance of a metamaterial element effectively tunes the complex polarizability of the equivalent dipole, shifting both the magnitude and phase with only one control knob \cite{jouvaud2013SRRantenna}; as a result, using metasurface elements to generate an arbitrary magnitude pattern without regard to the associated phase---or an arbitrary phase pattern without regard to the associated magnitude---can yield undesirable or unpredictable results. While it is possible to achieve independent control over both phase and magnitude through the combination of passive electric and magnetic metamaterial resonators, as has been demonstrated in \emph{Huygens} metasurfaces \cite{huygens}, the inclusion of two sets of resonators adds general complication as well as the possibility of mutual interactions between resonators \cite{smith_magnetoelectric}---a difficulty compounded for dynamically tunable metasurfaces that require bias and control circuitry.

Our goal here is to develop an optimal strategy for using individual resonant metamaterial elements as the radiating elements in a metasurface antenna. In Section \ref{MetasurfaceElementTuning} we first describe the polarizability model associated with metamaterial elements, with the essential properties of the waveguide-fed metasurface antenna and basic beam forming within the metasurface paradigm subsequently described in Section \ref{HolographicAntennas}. In Section \ref{GeneralizedMapping}, we introduce the general mapping concept with several examples, and in Section \ref{simulations} illustrate the benefits on beam steering performance, as well as validate the analytical model using numerical simulations. We provide some concluding remarks in Section \ref{sec:conclusion}.

\section{Polarizability Model}
\label{MetasurfaceElementTuning}

As is well-known in antenna engineering, an array of radiating elements can form arbitrary wave forms---limited only by diffraction effects---if there is complete control over both the amplitude and phase of the radiated field from each element \cite{phase_array_review}. To achieve such control, including a full $2\pi$ phase shift of an element, active components are typically introduced into the feed structure at each radiating node. The metasurface antenna, by contrast, is a passive device, in which control over element radiation is achieved by tuning the resonance of the magnetic polarizabilities of the metamaterial elements, each of which has the general Lorentzian form

\begin{equation}
\alpha_m(\omega) = \frac{F\omega^2}{\omega_0^2-\omega^2+j\Gamma \omega}.
\label{eq:lorentzian}
\end{equation}

\begin{figure}
\centering
\includegraphics[width=6cm,trim=1cm 0cm 1cm 0cm]{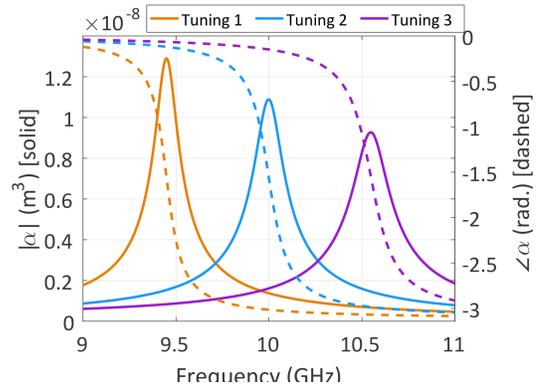}
\caption{Lorentzian polarizability for the idealized metamaterial element used to model a sample metasurface antenna. The three colors correspond to three different tuning states. The solid lines plot $|\alpha|$, while the dashed lines plot $\angle \alpha$}
\label{fig:alpha}
\end{figure}

\noindent Here, $F$ is a coupling factor, $\omega_0$ is the resonance frequency and $\Gamma$ is a damping factor. Although a wide range of phase shifts and amplitudes are possible by this approach, they are constrained by the Lorentzian form, as depicted in Fig. \ref{fig:alpha}. Note that the phase shift is restricted to the range $0$ to $-\pi$, with the amplitude falling to very low values at the extremes of the phase shift range. Eq. \ref{eq:lorentzian} can be used to obtain a convenient and compact relationship that expresses the constrained nature of the phase and amplitude of the polarizability, or

\begin{equation}
\alpha=F Q \sin \theta e^{-j\theta}.
\label{eq:lorentzian_constraint}
\end{equation}

\noindent where \(Q=\omega_0/\Gamma\) and \(\theta=\tan^{-1}(-\Gamma\omega/(\omega_0^2-\omega^2))\). Note that the amplitude of the polarizability is specified once the phase has been specified. It is clear from these equations that the magnitude of the polarizability can be modified by changing the coupling factor $F$ or the quality factor $Q$. Changing the resonance frequency, $\omega_0$ produces both a shift in phase as well as in magnitude. While adjusting the coupling or Q factor are potential methods of adjusting the polarizability of an element, these parameters tend to be associated with the physical geometry and intrinsic properties of the metamaterial element and are thus less convenient as a control mechanism. The resonance frequency, by contrast, relates to the electromagnetic environment and can be tuned by changing the capacitance (or inductance) of the element. In the analysis presented here, we thus consider modifying the polarizability of a metamaterial element by changing its resonance frequency, without further discussion of any specific physical mechanism.

Once a metasurface antenna architecture has been defined, a polarizability distribution must be determined for the metamaterial elements that will produce a desired radiation pattern. A standard approach is that of holography, in which the fields corresponding to a desired radiation pattern are backpropagated to the aperture and interfered with the incident feed wave, resulting in the required aperture field distribution. The aperture field distribution is then equated to a fictitious surface current distribution, which can finally be reduced to a discrete set of polarizability values spaced periodically over the aperture. The polarizabilities arrived at in this manner are unconstrained and represent the \emph{ideal} holographic solution. Since the range of phases and magnitudes required for the ideal solution are generally not available from the Lorentzian-constrained metamaterial resonators that obey Eq. \ref{eq:lorentzian_constraint}, a mapping must be found that relates each of the ideal polarizability values to the best choice from the available polarizability values. Different mapping schemes will translate to different antenna performance characteristics. 

A straightforward approach to achieving a desired radiation pattern would be to apply numerical optimization methods, starting perhaps from an approximation to the holographic solution. Such methods are frequently used, and can become necessary when the mutual coupling between elements is strong \cite{mesa}. For strongly coupled elements, the incident wave can be significantly modified by the interactions, thus requiring a self-consistent solution to an inverse problem. Numerical optimization can achieve this goal, but generally requires numerous iterations and is inherently computationally expensive. As an alternative to numerical optimization, specifically in the weak coupling limit where the incident wave is unperturbed, we introduce here a continuous family of mapping schemes that are simply visualized by viewing the ideal and constrained polarizability values in the complex plane. By adjusting a single parameter for this analytic mapping, we illustrate how the performance of the aperture can be optimized, leading to the generation of desired radiation patterns and excellent beam forming performance. This mapping technique is broadly applicable to metasurface antennas, not only the waveguide-fed variety analyzed here. In many scenarios, one of the mapping schemes presented here may constitute sufficient optimization, obviating further numerical optimization steps.

\section{Waveguide-Fed Metasurface Antenna}
\label{HolographicAntennas}
The waveguide-fed metasurface antenna architecture considered here consists of a waveguide in which numerous metamaterial elements are patterned along one of the conducting surfaces \cite{holloway2009metasurface,brady2009CSholo,PhysRevApplied.8.054048}. The guided wave excites the metamaterial elements, which couple energy from the waveguide to radiation. In analogy with conventional holography, the guided wave can be considered a reference wave, which interferes with the array of metasurface elements to form the desired radiation pattern (object wave). If the phase and magnitudes of the metasurface elements are unconstrained, then the collection of elements could be considered a hologram in the classical sense. In this work we consider only metasurface antennas in which the guided wave is restricted to propagation along one direction, which we refer to as one-dimensional (1D). We choose a coordinate system such that the normal to the surface is in the $\hat{z}$ direction, while propagation along the waveguide is in the $\hat{y}$ direction. For clarity, in the ensuing analysis we assume all metamaterial elements are oriented such that their effective magnetic dipole moments lie along the same direction. This limitation is easily removed using the tensor formalism provided in the appendix, so that metasurface antennas with polarization control can be considered within the presented framework.

The scattered field of any object or antenna can be described by the object's far-field pattern \cite{balanis2005antenna,yaghjian2005antennas,yaghjian1984OEWG}. For the application of beam forming and steering, we define as ideal an antenna that only radiates in a single direction; in this situation, the field produced by the antenna is a plane wave, so that 
\begin{equation}
\label{eq:planewave}
\textbf{E}(\textbf{r})=\textbf{E}_0e^{-j\textbf{k}_b\cdot\textbf{r}},
\end{equation}


\noindent where \(\textbf{E}_0\) is the polarization of the electric field and \(\textbf{k}_b\) is the desired direction of the antenna beam. We assume here that the electric fields lie in the $\hat{y}$-$\hat{z}$ plane. Eq. \ref{eq:planewave} represents the object or output wave. If the antenna were to extend infinitely in a plane \(S\) with unit-normal vector \(\hat{\textbf{n}}\), then the set of currents that would produce this far-field pattern are a set of magnetic surface currents over a perfectly electric conducting (PEC) ground plane, given for points \(\textbf{r}\in S\) as

\begin{equation}\label{eq:FundamentalDesignEqn1d}
\textbf{K}_M=\textbf{E}\times\hat{\textbf{n}}=E_{0y} e^{-j\textbf{k}_b\cdot\textbf{r}}\hat{x}.
\end{equation}

\noindent Eq. \ref{eq:FundamentalDesignEqn1d} can be proven using Schelkunoff's equivalence principles \cite{Collin1991,harrington2015time}. One of Schelkunoff's equivalence principles states that if the electric and magnetic fields are known on the boundary \(S\) of a domain \(V\) that contains no sources, then the field inside \(V\) is equal to the field that would be radiated by a magnetic current \(\textbf{K}_M=\textbf{E}\times\hat{\textbf{n}}\) that lies on \(S\), and is backed by a PEC just behind \(S\).  However, Schelkunoff's equivalence principle might also be stated backwards: if one desires a particular electric field \(E\) in a volume \(V\) that is bounded by a surface \(S\), then placing a magnetic surface current of \(\textbf{K}_M=\textbf{E}\times\hat{\textbf{n}}\) along with a PEC along \(S\) will achieve the electric field \(E\), provided that \(E\) is a solution to the wave equation.  If we consider the volume \(V\) to be the upper half-space with \(S\) being the plane where \(z=0\), then the inverse of Schelkunoff's equivalence principle dictates that the magnetic current distribution in Eq. \ref{eq:FundamentalDesignEqn1d} will radiate like a plane wave in the direction \(\textbf{k}_b\).

In the context of the metasurface antenna, the goal is to produce a required surface current \(\textbf{K}_M\) using waveguide slots or complimentary metamaterial elements (such as complementary electric inductive-capacitive resonators, or cELCs) \cite{Landy,stevenson2016mtenna,lipworth2013JOSA,maci2011metasurfing,mario1767metasurfing,minatti2012metasurface}.  These elements are known to radiate like magnetic dipoles and can provide the required magnetic surface current to produce an ideal plane-wave beam in the limit that the antenna is infinitely large \cite{pulido2016dda,Landy, bowen2012DDA}. Assuming the feed magnetic field and magnetic current are both oriented in the $\hat{x}$ direction, the magnetic surface current can be written as a magnetic surface susceptibility \(\chi_M\) multiplied by the magnetic field of the feed wave \(H_f \hat{x}\):

\begin{equation}\label{eq:ProducingSurfaceCurrent1d}
K_M(\textbf{r})=-j\omega \mu_0 \chi_M(\textbf{r}) H_f(\textbf{r}).
\end{equation}

The surface susceptibility is defined as the magnetic dipole moment generated per unit area on the surface of the antenna \cite{Jackson,Collin1991,karamanos2012polarizability}. To achieve a well-formed beam, the surface current needs to be set equal to \(\textbf{E}_0\times\hat{\textbf{n}}e^{-j\textbf{k}_b\cdot\textbf{r}}\), which can be done by carefully designing the surface susceptibility distribution \(\chi_M\) using the appropriate modulation pattern \cite{holloway2011characterizing}.    

For waveguide-fed metasurface antenna designs, the coupling of each metamaterial element to the waveguide mode is desired to be weak, allowing the waveguide mode to fill as much of the aperture as possible. For very weakly coupled elements, the scattering back into the waveguide can be considered negligible and the magnetic field of the feed wave unperturbed. Under these assumptions, the surface susceptibility at position \(\textbf{r}_i\) is related to the magnetic polarizability \(\alpha_i\) of the element at position \(\textbf{r}_i\) by \(\alpha_i=\Lambda^2\chi(\textbf{r}_i)\). Here, $\Lambda$ is the spacing between metamaterial elements. Using this relationship, together with Eqs. \ref{eq:FundamentalDesignEqn1d} and \ref{eq:ProducingSurfaceCurrent1d}, the required polarizabilities of the dipoles are

\begin{equation}
\label{eq:PlanarWgIdealModulation1d}
\alpha(\textbf{r}_i)\&=\frac{j\Lambda^2}{Z_0 k} E_{0y} \left(\frac{e^{-j\textbf{k}_b\cdot\textbf{r}_i}}{H_f}\right). 
\end{equation}

\noindent Here we have used the dispersion relation $\omega=c k$, where $c^{-2}=\epsilon_0 \mu_0$ and $Z_0^{2}=\mu_0/\epsilon_0$.

At this point, a particular form of the feed wave needs to be chosen. For a 1D metasurface antenna, the feed structure is a linear waveguide that has a single propagating mode with magnetic field on the upper surface of the waveguide \(H_f=h_fe^{-j\beta y}\), where \(\beta\) is the propagation constant and the waveguide is assumed to lie in the \(y\)-direction. In this case, the requirement on the polarizabilities to produce an ideal beam pattern is

\begin{equation}
\alpha(\textbf{r}_i)\&=\frac{j\Lambda^2}{Z_0 k}\left(\frac{E_{0y}}{h_f}\right)e^{-j(\textbf{k}_b-\beta\hat{\textbf{y}})\cdot\textbf{r}_i} \label{eq:LinearWgIdealModulation1d}
\end{equation}

\noindent Note that the term \(\left(\frac{E_{0y}}{h_f}\right)\equiv Z_{ant}\) has units of impedance and defines the amplitude of the electric field of the radiated wave relative to the amplitude of the magnetic field of the feed wave.  

As the index \(i\) runs through various metamaterial element positions \(\textbf{r}_i\), the ideal modulation pattern will be proportional to \(e^{-j(\textbf{k}_b-\beta\hat{\textbf{y}})\cdot\textbf{r}_i}\), which traces out a unit circle in the complex plane. This modulation compensates for the phase accumulation of the guided wave while applying the pattern required to form a beam in the $\textbf{k}_b$ direction. Hence, in an ideal situation, all of the elements would be excited at an equal level, but with different scattering phases dictated by the modulation pattern.

\begin{figure}[h]
\centering
\includegraphics[width=0.45\textwidth]{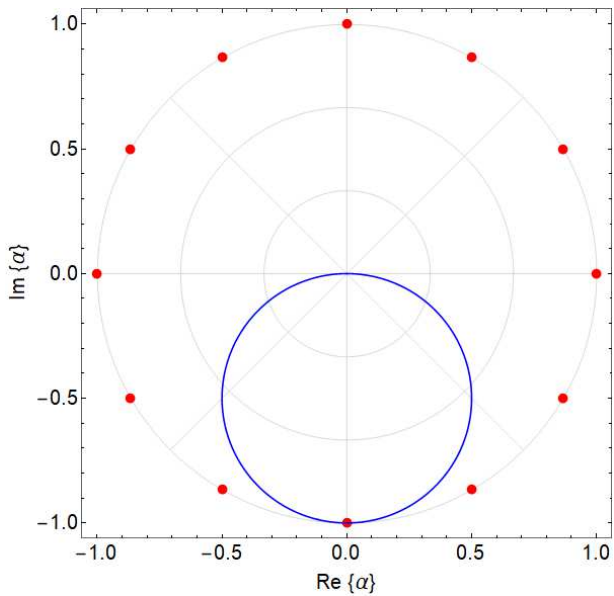}
\caption{Desired polarizabilities (shown as red dots) required to achieve a directive beam and the Lorentzian constrained region (blue circle) achievable by tuning metasurface elements. The axes are scaled in units of $F Q$.}\label{fig:problem}
\end{figure}

A plot of the \emph{available} polarizabilities (Eq. \ref{eq:lorentzian_constraint}) in the complex plane (Fig. \ref{fig:problem}) as a function of the resonance frequency \(\omega_0\) provides a visual representation of the range of polarizabilities that may be achieved by tuning the resonance frequency of a metamaterial cell. However, Eqs. (\ref{eq:PlanarWgIdealModulation1d}) and (\ref{eq:LinearWgIdealModulation1d}) prescribe the polarizabilities that are required for each metamaterial element in order for the antenna to radiate in a far-field pattern that most closely approximates an ideal plane wave. Unfortunately, the set of available polarizabilities does not overlap with the set of ideal polarizabilities.  

In Fig. \ref{fig:problem}, the real and imaginary parts of Eq. \ref{eq:lorentzian_constraint} are plotted as a blue curve in the complex plane that is parameterized by \(\omega_0\) for the choice of parameters  \(\Gamma=\omega/50\) and \(F=(\lambda/3)^3\), illustrating the range of achievable polarizabilities using metamaterial elements. The same plot also shows the required polarizabilities from Eq. \ref{eq:LinearWgIdealModulation1d} for a linear waveguide with metamaterial elements that are placed \(\Lambda=\lambda/5\) apart on the \(y\)-axis, and \(\beta=1.5k\) (red dots).  We note that values of polarizability that lie on the upper half of the complex plane are forbidden if the dipole is to be passive in the \(e^{j\omega t}\) time convention, regardless of whether or not the polarizability follows a Lorentzian resonance. Note also that we have selected the coupling parameter of the Lorentzian constrained elements so that the diameter of the circle indicating the Lorentzian constrained region is equal to the radius of the circle on which the ideal polarizabilities lie. This particular choice is arbitrary; the Lorentzian constrained region can be larger or smaller depending on the coupling parameter. We apply this particular choice here for consistency. In the limit that the reference wave is unperturbed, the choice of scaling makes no difference in the antenna directivity patterns that will be investigated later.


The Lorentzian constrained polarizability curve only tangentially approaches the ideal polarizability curve at a single point. A choice must thus be made that maps the ideal polarizabilities to the available Lorentizian-constrained polarizabilities, with the hope that such a choice will produce the desired radiation pattern with minimal error. 

Before addressing the mapping approach that is the central concept of this work, we note two simple strategies for optimizing polarizability distributions: \emph{phase holograms} and \emph{magnitude holograms}. In the case of a phase hologram, the modulation scheme can be derived in the perturbative approximations of holography\cite{goodman1996introduction} by interfering the complex conjugate of the reference (in this case guided) wave with the desired object wave, which yields the ideal phase of the polarizability for each metamaterial element. For the phase hologram, the magnitude would be assumed as constant. In the case of a magnitude hologram, the desired magnitude of the surface currents is assumed to be sinusoidally varying. Here, the magnitude of each element's polarizability is tuned to match the desired magnitude profile by taking the real part of the phase hologram, linearly mapping it to a positive real space. For the magnitude hologram, the phase would be considered as constant. 

While a true magnitude hologram could be implemented using attenuators or amplifiers, and a true phase hologram implemented using phase shifters, the passive metamaterial elements do not possess this degree of freedom. We have thus termed these two holographic approaches as \emph{Lorentzian-constrained} because they represent a particular choice of polarizability functions that are consistent with Eq. \ref{eq:lorentzian} and might be considered natural and straightforward. While both of these methods are able to generate directive beams in the far-field, both incur undesired side-effects. In the case of a phase hologram approach, the goal of achieving a desired phase profile can be met, but in doing so many elements with the correct phase will be incidentally set to have low or zero magnitude, resulting in suboptimal performance \cite{johnson2015sidelobe}. Likewise, in the case of a magnitude hologram, the goal of matching the magnitudes to the desired profile may be met, but in doing so, many elements may have different phases which can interfere to degrade performance in the far-field.

With the Lorentzian constraint it is impossible to vary either just the phase or just the magnitude, so that the resulting Lorentzian-constrained holograms are only inspired by their ideal counterparts. Thus, while the simple holographic approaches are conceptually appealing, they result in generally poor beam forming performance, thus motivating a design strategy that incorporates the linked phase and magnitude of the metamaterial elements.

\section{Generalized Mapping of Ideal to Available Polarizability Values}
\label{GeneralizedMapping}
Aside from the Lorentzian-constrained magnitude hologram, all of the modulation strategies considered here can be derived graphically in the complex plane by first mapping the ideal polarizability values to a single point along the negative imaginary $\alpha_m$ axis. For this and subsequent sections, we rewrite Eqs. \ref{eq:LinearWgIdealModulation1d} for the ideal polarizabilities as

\begin{equation}
\label{eq:simplifiedPolarizability}
    \alpha=x+j y=A e^{j \Psi},    
\end{equation}
 
\noindent where $\Psi=-\textbf{k}_b \cdot \textbf{r}_i+\beta \hat{y} \cdot \textbf{r}_i -\pi/2$. Taking the real constant $A=1$, the polarizability values in the complex plane thus occur at positions ($\cos{\Psi},\sin{\Psi}$). The equation for a line extending from one of these polarizability values to a point ($0,y_0)$ on the negative $\alpha_m$ axis thus has the form

\begin{equation} \label{eq:mapY}
    y=m x + y_0
\end{equation}

\noindent where

\begin{equation} \label{eq:slope}
m= \frac{\sin{\Psi}-y_0}{\cos{\Psi}}.
\end{equation}

\noindent The mapping for a given ideal value occurs where this line intersects the circle corresponding to the Lorentzian-constrained polarizability values. The equation for the Lorentzian-constrained curve is a circle of the form

\begin{equation} \label{eq:mapX}
x^2+ ( y+r )^2=r^2.
\end{equation}

\noindent where $r$ is the radius of the Lorentzian-constrained circle. As noted above, for this analysis we set $r=0.5$ since the value does not matter when the feed wave is unperturbed. When there is significant coupling of energy from the waveguide to radiation, the value of $r$ does matter and can be used as a second parameter to optimize the beam performance. The impact of $r$ will be presented in a future study.

\begin{figure}[h]
\centering
\includegraphics[width=0.45\textwidth]{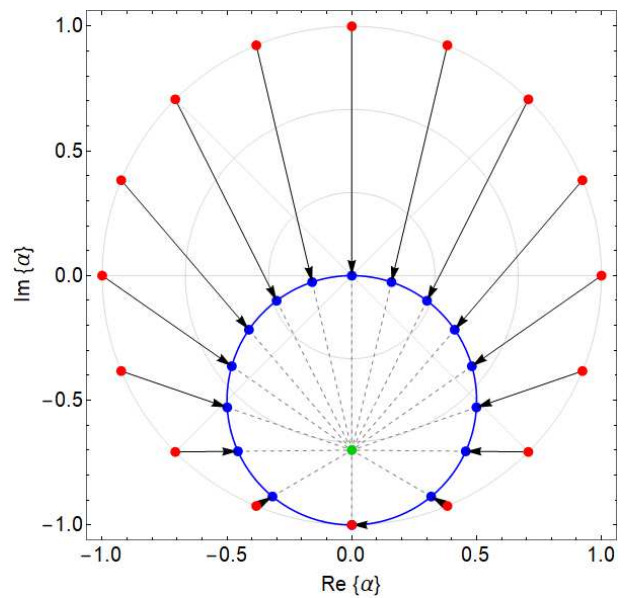}
\caption{Mapping between ideal polarizabilities (shown as red dots) required to achieve a directive beam and the available Lorentzian-constrained values (blue circle) for $y_0=-0.7$.}\label{fig:exampleMapping}
\end{figure}

Eqs. (\ref{eq:mapY}, \ref{eq:mapX}) can be solved simultaneously to yield

\begin{equation} \label{eq:mapSolution}
    x=\frac{-2 m y'\pm \sqrt{4 m^2 y'^2-4(m^2+1)(y'^2-r^2)}}{2(m^2+1)},
\end{equation}

\noindent where

\begin{equation}
    y'=y_0+r
\end{equation}

\noindent with $y$ determined from Eq. \ref{eq:mapY}. This process produces a distinct mapping of the ideal polarizability values to the Lorentzian-constrained values, which are the intersections of the set of lines with the Lorentzian-constrained circle. Each value of $y_0$ corresponds to a distinct mapping, as illustrated in Fig. \ref{fig:exampleMapping} for a value of $y_0=-0.7$.

Having defined the basic mapping, we now consider several specific examples and study the impact on beam forming.

\subsection{Lorentzian-Constrained Phase Hologram} 
Consider first the case in which $y_0=0$. The two possible signs of the square root in Eq. \ref{eq:mapSolution} can be evaluated as

\begin{equation}
\begin{aligned} 
    x &=  0 & 0<\Psi<\pi\\ 
    x &=  -\frac{1}{m+\frac{1}{m}} & \pi < \Psi < 2\pi
\end{aligned}
\end{equation}

\noindent Since $m=\tan{\Psi}$, we arrive at the following mapping:

\begin{equation}
\begin{aligned} 
    \alpha &=  0 & 0<\Psi<\pi\\ 
    \alpha &=  -\sin{\Psi}e^{j \Psi} & \pi < \Psi < 2\pi
\end{aligned}
\end{equation}

The resulting mapping corresponds to a simple phase hologram, since the phases of the Lorentzian-constrained polarizabilities correspond exactly to those of the ideal polarizabilities. Note that all of the phases in the upper half plane are mapped to zero and are therefore unused. It can thus be expected that the resulting beam forming performance of the aperture will be sub-optimal, with so many unused radiators. This mapping with $y_0=0$ is illustrated in Fig. \ref{fig:modulation_mapping} (top). As discussed in the previous section, the magnitudes of the non-zero polarizabilities are modulated by the sine term, as necessitated by the Lorentzian constraint.

\subsection{Lorentzian-Constrained Amplitude Hologram} 
Following strategies used in holography, one can consider a modulation scheme in which only the amplitudes of polarizabilities are modified. While this mapping deviates from the family of mappings that are the focus of this paper, we include the Lorentzian-constrained amplitude hologram for completeness and later comparisons. The real part of the ideal polarizability, Eq. \ref{eq:simplifiedPolarizability}, is $\alpha^i=A \cos{\Psi}$, which would form the basis for an amplitude hologram. This quantity has both positive and negative values, so to ensure only positive real values are used, we consider the modulation $\alpha^i=A(1+ \cos{\Psi})/2$. Such a distribution of polarizaibility values is possible by changing only the Q or the coupling factor; however, we are focused here on modifying the resonance frequency of the elements to obtain a set of polarizability values, such that the phase is inherently changed as a function of magnitude. In fact, given the form of the Lorentzian constrained polarizability, Eq. \ref{eq:lorentzian_constraint}, we must have

\begin{equation}\label{eq:magnitude_mapping}
|\alpha^i|=\sin{\theta}= \frac{1- \cos{\Psi}}{2}
\end{equation}

\begin{figure}[h]
\centering
\includegraphics[width=0.4\textwidth]{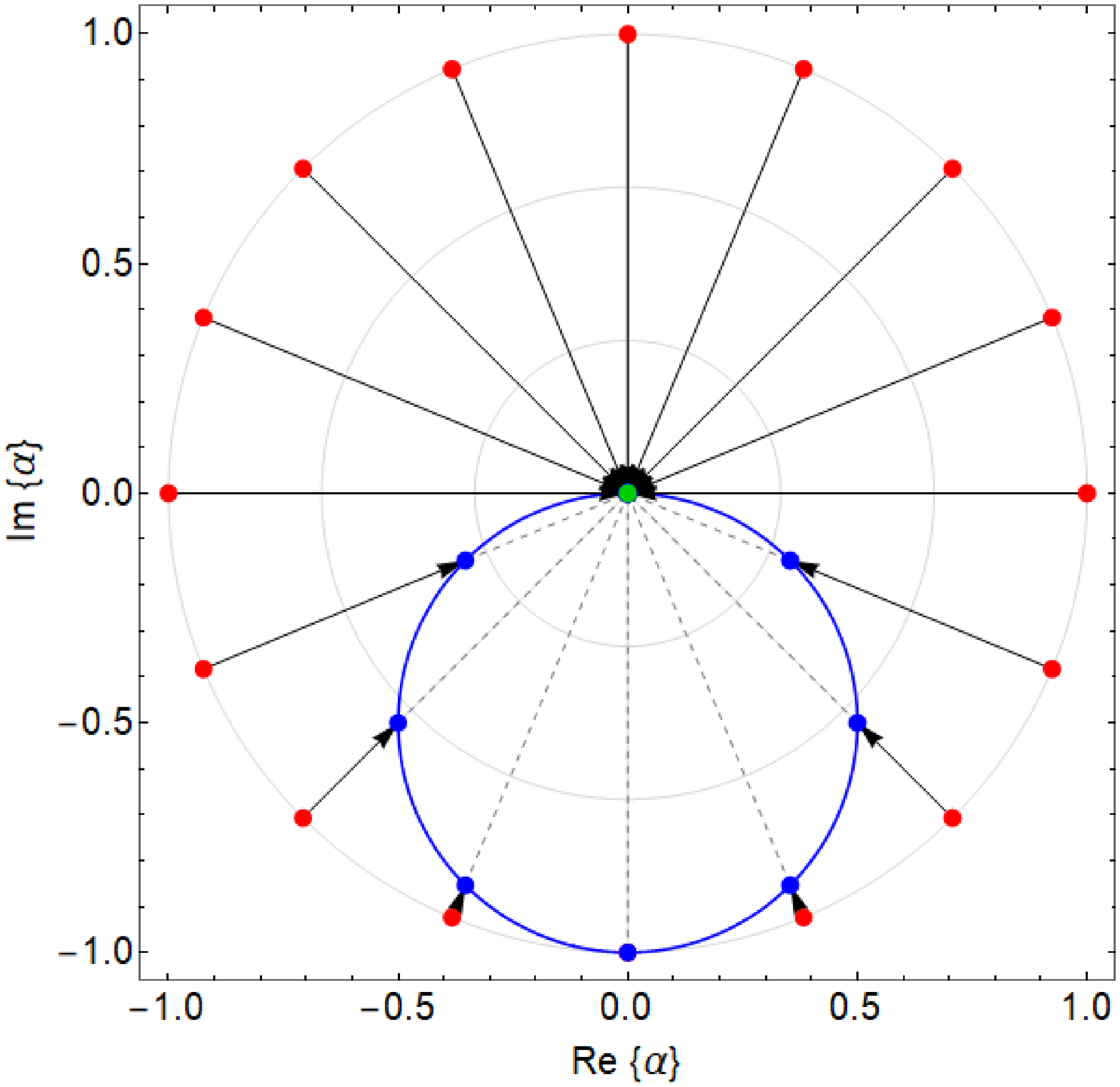}

\includegraphics[width=0.4\textwidth]{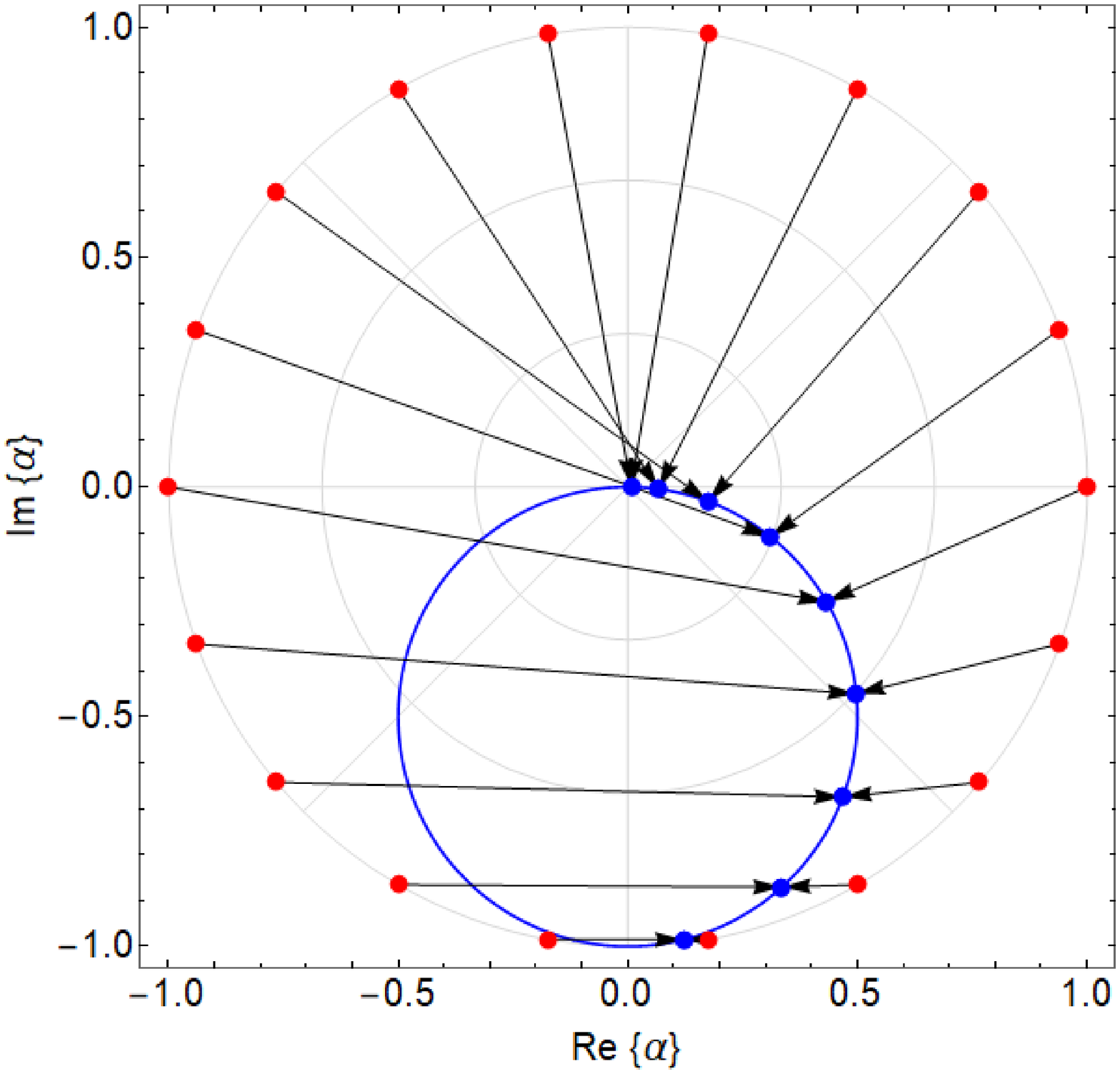}

\caption{The mapping between the ideal polarizabilities (red dots) and those corresponding to a simple phase hologram ($y_0=0$, top) and a simple magnitude hologram (bottom). The ideal polarizabilities (red dots) are plotted in the complex plane along with the constrained Lorentzian polarizabilities (blue curve). The black arrows represent the mapping between the ideal solution and the solution that is achievable for resonant elements.}\label{fig:modulation_mapping}

\end{figure}

\noindent Eq. \ref{eq:magnitude_mapping} provides the phase $\theta$ of the Lorentzian constrained elements as a function of $\Psi$, allowing the magnitude mapping to be represented as in Fig. \ref{fig:modulation_mapping} (bottom).

This somewhat artificial construct modifies the modulation circle of desired polarizabilities in the complex plane by mapping it onto a set of positive, real numbers.  However, once this is done, it is possible to ignore the phase of the desired modulation pattern, and hence select tuning states of the elements by finding the amplitude of the set of tunable polarizabilities from the Lorentzian elements, which are given by taking the magnitude of Eq. \ref{eq:lorentzian_constraint}. A \emph{Lorentzian-constrained amplitude hologram} is therefore achieved by finding the points in the curve of tunable polarizabilities that are closest in amplitude to the amplitude prescribed in Eq. \ref{eq:simplifiedPolarizability}. As opposed to the phase hologram, which makes only one approximation by ignoring the magnitude, the amplitude hologram therefore has two steps in approximation. The first approximation is that it ignores the phase of the response of the elements, and the second approximation is that it artificially maps the ideal modulation pattern equation to the positive real axis. Clearly, the phase varies along with the magnitude modulation, so that the resulting modulation pattern is more complicated and generally much less predictable than would be a true magnitude only hologram.

\subsection{Array Factor Optimized Modulation} 

\begin{figure}[h]
\centering
\includegraphics[width=0.4\textwidth]{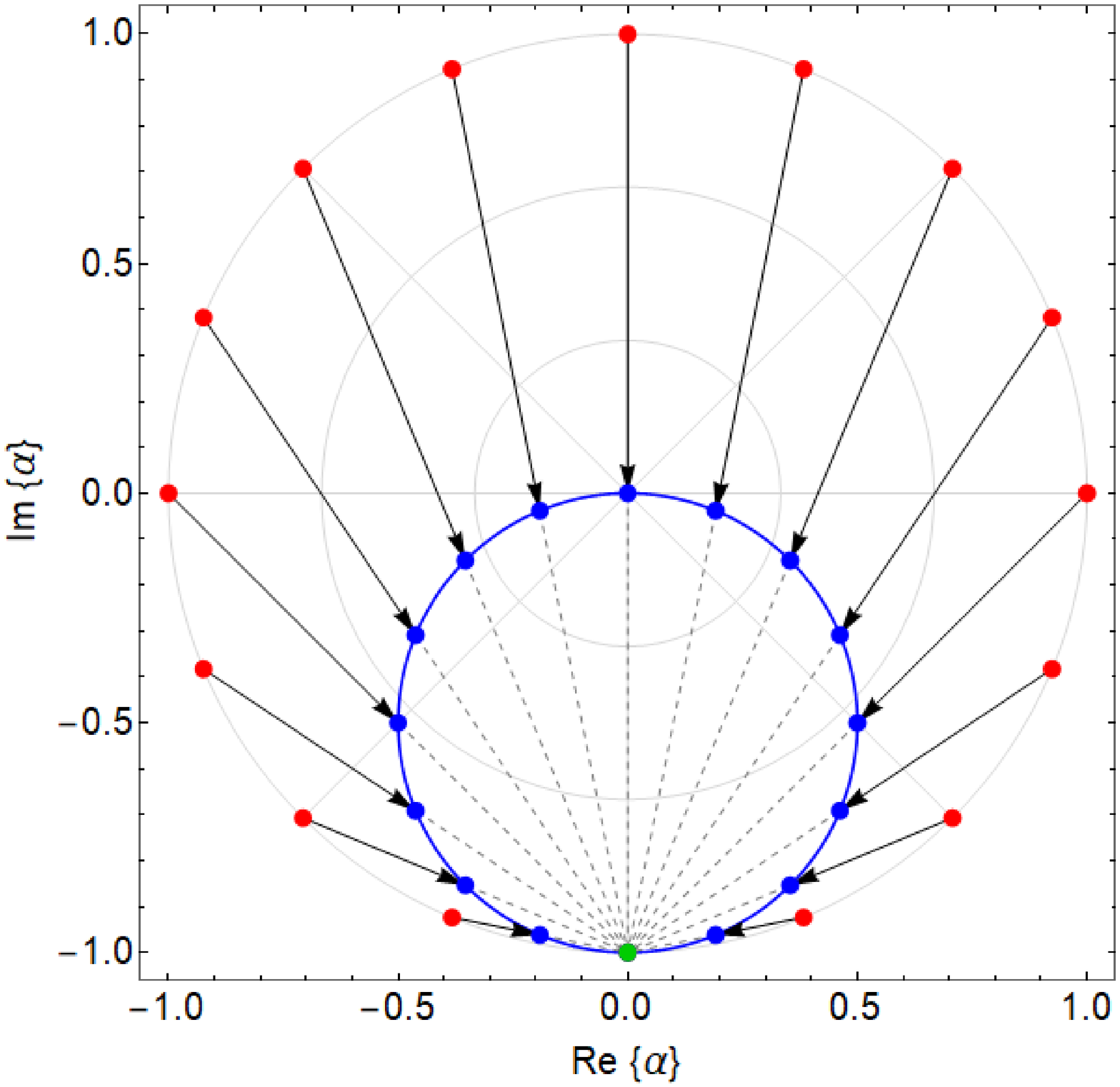}

\includegraphics[width=0.4\textwidth]{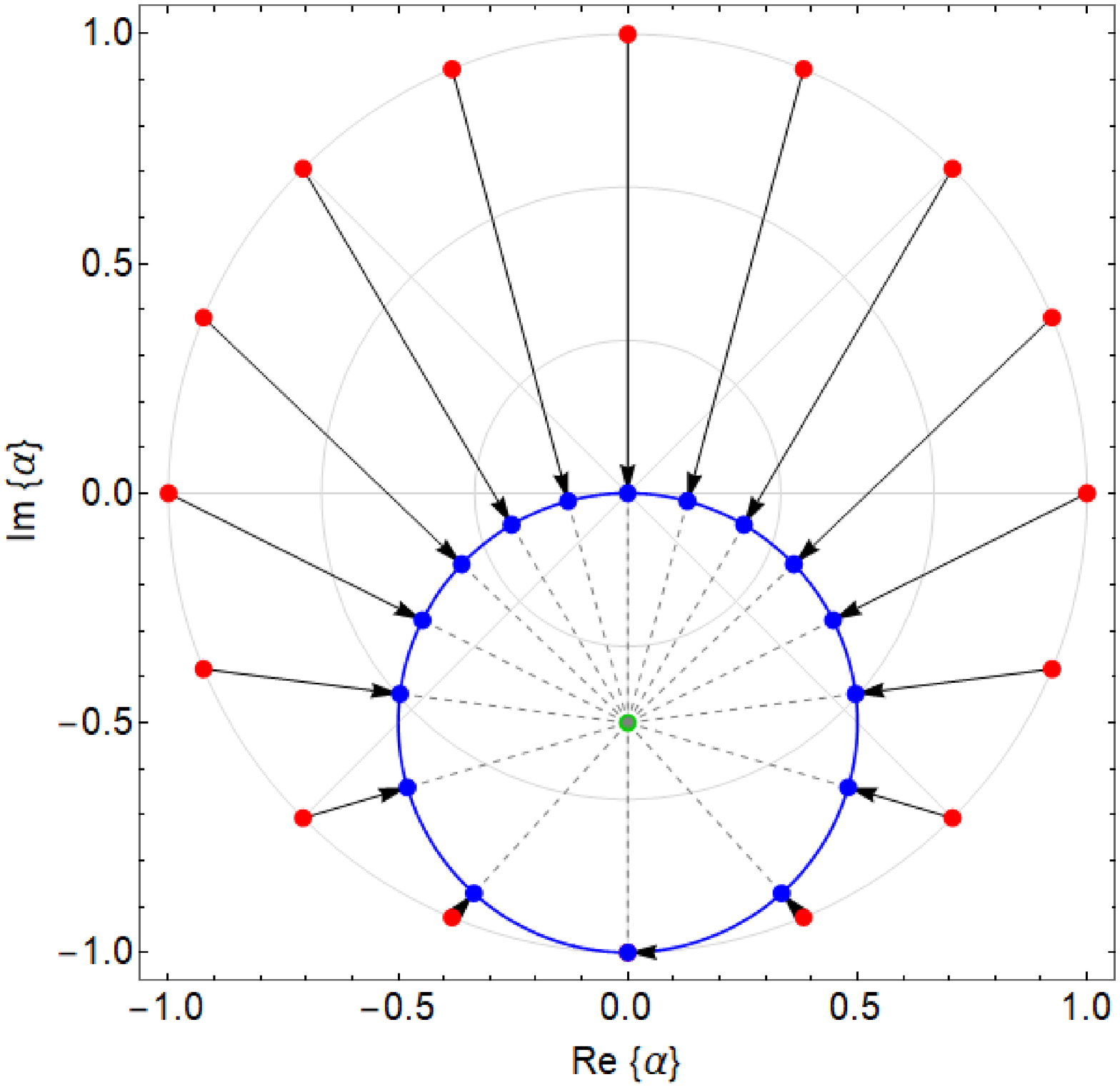}
\caption{Mappings corresponding to the case of $y_0=-1$ (top) and the Euclidean mapping scheme ($y_0=-0.5$, bottom).}\label{fig:modulation_mapping_euclidean}
\end{figure}

At the extreme value of $y_0=0$, half of the ideal polarizability values are mapped to zero, with many more mapped to regions of the Lorentzian-constrained curve where the amplitude is small. As the focal point is moved down the $y$-axis, the ideal polarizability values are increasingly mapped to regions of the Lorentzian-constrained curve where the magnitude of the polarizability is larger, such that it might be expected these mappings will have improved efficiency and performance. In this context, it is useful to consider the extreme value of $y_0=-1$. Using

\begin{equation}
    m=\frac{\sin{\Psi}+1}{\cos{\Psi}}
\end{equation}

\noindent with Eq. \ref{eq:mapSolution}, we find

\begin{equation}
    x=\frac{1}{m+\frac{1}{m}}=\frac{\cos{\Psi}}{2}
\end{equation}

\noindent and

\begin{equation}
    y=\frac{\sin{\Psi}-1}{2},
\end{equation}

\noindent which can be combined together to yield the polarizability

\begin{equation}
\label{eq:lorentzianPolarizability}
    \alpha=\frac{1}{2 j}(e^{j \Psi'}-1)
\end{equation}

\noindent where $\Psi'=\Psi+\pi/2$. This mapping is shown in Fig. \ref{fig:modulation_mapping_euclidean} (top). (Note that the ideal and Lorentzian constrained phases are also rotated by $-90^\circ$ due to the $j$ term in the denominator.) From the figure, it can be observed that, unlike the mappings for other values of $y_0$, the $y_0=-1$ mapping results in a uniform density of values over the Lorentzian-constrained circle and the largest number of values having non-negligible magnitudes.

The simple analytic form for the $y_0=-1$ mapping is particularly convenient when used in conjunction with the array factor for the 1D antenna, as noted in \cite{PhysRevApplied.8.054048}. For beam forming in the plane of the antenna, the array factor has the form

\begin{equation}\label{eq:array_factor}
    AF(\phi)=\sum_{i=1}^N \alpha e^{-j (\beta \hat{\textbf{y}}-\textbf{k}_o)\cdot \textbf{r}_i}
\end{equation}

\noindent where $k_o$ is the wave vector in the direction of observation ($\textbf{k}_o \cdot \textbf{r}_i=k_0 r_i \sin{\phi}$). Inserting the mapped polarizability values (Eq. \ref{eq:lorentzianPolarizability} into the array factor yields

\begin{equation}
    AF(\phi) = \sum_{i=1}^N \frac{1}{2j}(e^{j (\beta \hat{y}-\textbf{k}_b)\cdot \textbf{r}_i}-1)e^{-j(\beta \hat{\textbf{y}}-\textbf{k}_o)\cdot \textbf{r}_i},
\end{equation}

\noindent or

\begin{equation}
    AF(\phi) = \sum_{i=1}^N \frac{1}{2j}(e^{j (\textbf{k}_o-\textbf{k}_b)\cdot \textbf{r}_i}-e^{j (\beta \hat{y}-\textbf{k}_b)\cdot \textbf{r}_i}).
\end{equation}

\noindent the first term corresponds to the desired steered beam, with the second term corresponding to a possible diffracted beam. If the aperture is sampled at points that are spaced considerably closer than the free space wavelength, this term will not produce a radiating diffracted beam and the antenna will produce the single desired lobe.

\begin{figure}
\centering
\includegraphics[width=4.5cm,trim=1cm 0cm 1cm 0cm]{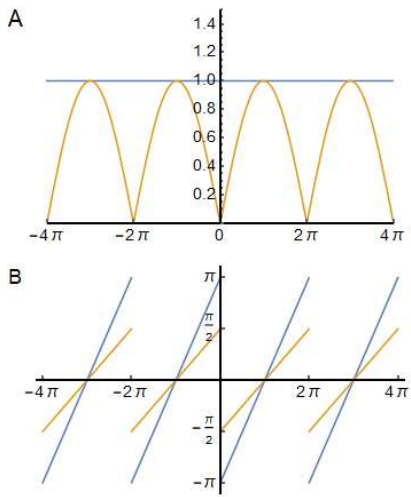}
\caption{(A) Amplitude and (B) phase of the Lorentzian-constrained polarizability (orange). The ideal amplitude and phase are shown for comparison as the blue curves. Note that the ideal phase has been shifted by $\pi$ to facilitate a direct comparison.}
\label{fig:lorentzian_constrained_amp_phase}
\end{figure}

It is useful for comparison purposes to plot the amplitude and phase of the mapping implied by Eq. \ref{eq:lorentzianPolarizability} versus $\Psi$, as shown in Fig. \ref{fig:lorentzian_constrained_amp_phase}. It can be seen that, like the ideal phase, the Lorentzian-constrained phase is linear, with a slope half that of the ideal (and shifted by $\pi$). While the amplitude varies, the variation combined with the modified phase advance results in a set of polarizabilities that produce the desired steered beam. Given that typically the phase of the wave in the aperture is dominant over its magnitude in terms of determining the far-field radiation pattern, the linear phase advance produced by this mapping might be nearly optimal among this set of mappings---a hypothesis reinforced by the array factor analysis and numerical simulations shown later.

While it may be tempting to consider this modulation scheme as \emph{optimal}, it should be remembered that the scheme has been derived with the assumption that the feed wave is left unperturbed. In a realistic scenario, the feed wave will decay as it propagates along the waveguide (due to radiative and other losses); this modulation scheme thus serves as a possible initial distribution for further optimization when coupling is significant.

\subsection{Euclidean Modulation}
\label{EuclideanModulation}
Another interesting mapping strategy occurs for the value $y_0=-0.5$, which coincides with the center of the Lorentzian-constrained circle. This case, plotted in Fig. \ref{fig:modulation_mapping_euclidean}(bottom), has the property that every line is normal to the Lorentzian-constrained circle and thus minimizes the Euclidean distance from the corresponding point on the ideal polarizability circle. We refer to this mapping as \emph{Euclidean modulation}. An alternative expression of this mapping is defined by the resonance frequency \(\omega_{0i}\) for the \(i^{th}\) metamaterial element located at position \(\textbf{r}_i\) such that the distance
\begin{equation}
\label{eq:euclideanModulationExpression}
d=|\alpha_L(\omega_{0i})-\alpha_D(\textbf{r}_i)|
\end{equation}
is a global minimum. Here, $\alpha_D$ represents the desired or ideal polarizabilities, while $\alpha_L$ represents the Lorentzian-constrained polarizabilities. The $y_0=-0.5$ mapping provides a reasonable level of optimization of the polarizabilities and has the advantage that it can be applied directly through Eq. \ref{eq:euclideanModulationExpression}. This latter property can be advantageous when the ideal polarizability distribution is not a circle and the use of the analytic formulas may be inconvenient. Note in Fig. \ref{fig:modulation_mapping_euclidean} (bottom) that the density of accessible polarizability is greatest where the amplitude is near zero; however, the mapping still provides far greater performance than either the Lorentzian-constrained phase or magnitude holograms and is easy to apply.

For the $y_0=-0.5$ case, Eq. \ref{eq:mapX} and Eq. \ref{eq:mapY} can be solved analytically to find 

\begin{subequations}
\begin{align}
x=\frac{1}{2} \Big [ 1+\Big ( \frac{\sin{\Psi}-\frac{1}{2}}{\cos{\Psi}} \Big )^2 \Big ]^{-1/2} \label{eq:eucMapX} \\
y=\frac{1}{2} \Big [ \Big ( \frac{\sin{\Psi}-\frac{1}{2}}{\cos{\Psi}} \Big ) \Big [ 1+\Big ( \frac{\sin{\Psi}-\frac{1}{2}}{\cos{\Psi}} \Big ) \Big ]^{-1/2} - 1\Big ]. \label{eq:eucMapY}
\end{align}
\end{subequations}

\begin{figure}
\centering
\includegraphics[width=4.5cm,trim=1cm 0cm 1cm 0cm]{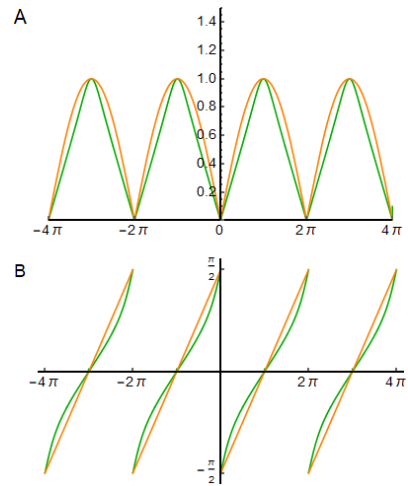}
\caption{(A) Amplitude and (B) phase of the Euclidean-constrained polarizability (green). The Lorentzian-constrained amplitude and phase are shown for comparison as the orange curves.}
\label{fig:euclidean_amp_phase}
\end{figure}

The amplitude and phase for the Euclidean modulation scheme ($y_0=-0.5$) are shown in Fig. \ref{fig:euclidean_amp_phase}, where they are compared with those obtained for the simple Lorentzian-constrained case ($y_0=-1.0$). The agreement between the two methods is striking, the major difference being that the phase shift is linear for the Lorentzian-constrained case, while the amplitude is nearly linear for the Euclidean modulation case. The close agreement suggests that Euclidean modulation also results in a fairly optimized beam-forming antenna. 

\section{Far-Field Performance}
\label{simulations}
To examine the far-field performance of a metasurface antenna using the various methods described above to determine the polarizability distribution, we perform array factor calculations using Eq. \ref{eq:array_factor}. The metasurface is assumed to be aligned along the \(y\)-axis with 64 radiating metamaterial elements that are spaced a distance of \(\Lambda=\lambda/4\), where \(\lambda\) is the free space wavelength. The driving frequency is assumed to be 10 GHz, so that the total aperture size is 16 $\lambda$. The guide index is assumed to be $n_g=1.6$.

\begin{figure}
\centering
\includegraphics[width=7cm,trim=1cm 0cm 1cm 0cm]{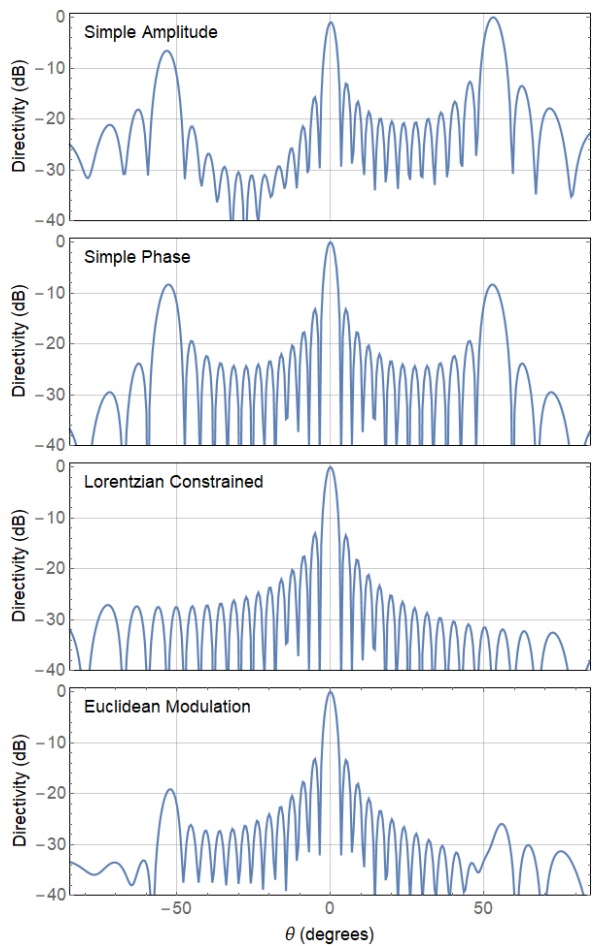}
\caption{Far-field radiation patterns for broadside beams optimized using the Lorentzian-constrained amplitude hologram, the Lorentzian-constrained phase hologram ($y_0=0$), the $y_0=-1$ mapping, and the $y_0=-0.5$ mapping (Euclidean modulation).}
\label{fig:broadside_beams}
\end{figure}

As described above, the excitation field is assumed to be a guided wave with a magnetic field that varies along the guide as $\textbf{H}_f(y)=h_f e^{-j\beta y}\hat{x}$. The guided wave is assumed to be unperturbed by the scattering of the elements in the waveguide, so that this study only examines the inherent benefits of each modulation technique, without including the effect of the attentuation of the feed wave. By exciting each metamaterial element with the guided wave, a magnetic dipole moment is induced as a function of $y$ proportional to the polarizabilty as determined by the given modulation technique. Thus, each dipole moment is given by \(\textbf{m}_i=\alpha(\omega_{0,i})\textbf{H}_f(y_i)\), where \(\omega_{0,i}\) is the tuned resonance frequency of the \(i^{th}\) unit cell, as prescribed by the chosen modulation technique. The collection of dipole moments can be propagated into the far field by summing the fields from each of the radiating dipoles. Since we are interested in the far-field behavior, we plot only the directivity using the array factor of Eq. \ref{eq:array_factor}. Sample radiation patterns for the various modulation schemes are shown in Fig. \ref{fig:broadside_beams}. 

\begin{figure}
\centering
\includegraphics[width=7cm,trim=1cm 0cm 1cm 0cm]{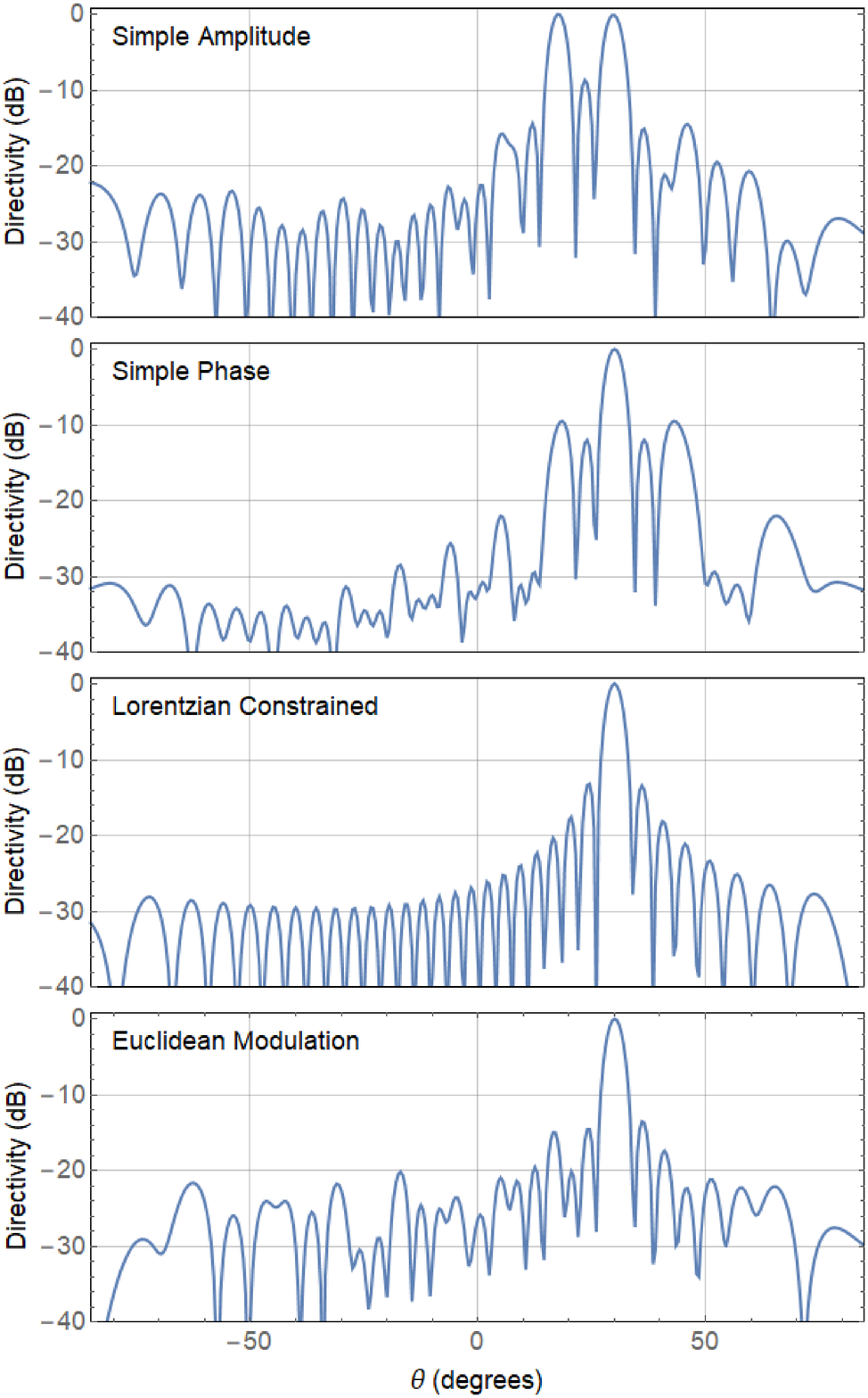}
\caption{Far-field radiation patterns for a beam steered to 30 degrees, optimized using the Lorentzian-constrained amplitude hologram, the Lorentzian-constrained phase hologram ($y_0=0$), the $y_0=-1$ mapping, and the $y_0=-0.5$ mapping (Euclidean modulation).}
\label{fig:steered_beams}
\end{figure}

\subsection{Directivity Comparisons}

A comparison of the four modulation schemes discussed is presented in Figs. \ref{fig:broadside_beams} and \ref{fig:steered_beams}. As would be expected, the simple phase and amplitude holograms display notably worse beam forming performance than the Lorentzian-constrained or Euclidean modulations schemes. In particular, grating lobes are prominent in the directivity patterns of the simple phase and amplitude modulation schemes, sometimes as large as the main lobe (as in Fig. \ref{fig:steered_beams}). In the case of the simple phase modulation, grating lobes are very likely due to the effective grating introduced by periodically switching all of the elements off. For the simple amplitude modulation, the arbitrary phase shift introduced by the mapping results in generally haphazard behavior of the directivity pattern, making the observed patterns irregular and relatively sensitive to even small changes in the parameters.

As expected, the Lorentzian-constrained modulation scheme ($y_0=-1$) shows the best performance, given the assumptions of the system---particularly that the waveguide mode is unperturbed. The pattern is essentially ideal, since no grating lobes are excited (as per the array factor). The Euclidean modulation scheme ($y_0=-0.5$) tends to perform well, even at steered angles, with minimal grating lobes. Because the phase advance is not quite linear for the Euclidean modulation scheme, there are numerous grating lobes with modest excitation; these lobes are relatively minor and can be minimized further with modest optimization.

The complex plane mappings can be readily applied in virtually any situation and thus provide a useful first guess for the polarizability distribution of a metasurface aperture antenna. If the interactions between elements and the feed wave are not significant, the distributions obtained by these mappings may, in fact, be all that is necessary in many cases.

\subsection{Numerical Simulations}
\begin{figure}[h!]
\centering
\includegraphics[width=7cm,trim=1cm 0cm 1cm 0cm]{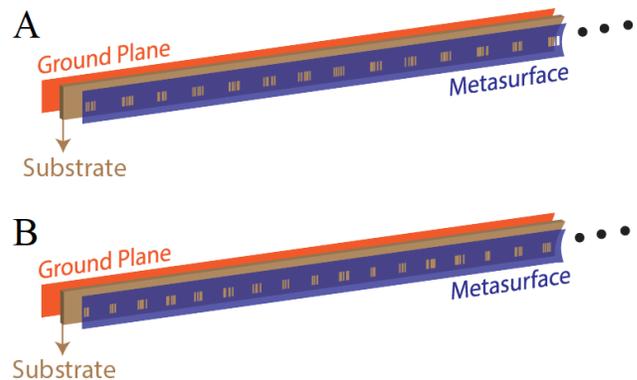}
\caption{Simulated metasurface antenna (a) broadside radiation (b) steered to 30 degrees. Only a portion of the metasurface antenna is shown to improve figure clarity.}
\label{fig:Metasurface_Antennas}
\end{figure}

To confirm the validity of the analytical far-field radiation patterns shown in Figs. \ref{fig:broadside_beams} and \ref{fig:steered_beams}, numerical simulations of the waveguide-fed holographic metasurface antenna were performed using the full-wave electromagnetic solver CST Microwave Studio. As illustrated in Fig. \ref{fig:Metasurface_Antennas}, the designed metasurface antenna is assumed to be constructed using printed circuit board (PCB) materials, similar to several waveguide-fed metasurface antennas recently demonstrated. The simulated structure consists of a ground plane, a dielectric substrate (assumed to be Rogers 4003C with a dielectric constant $\epsilon_r$=3.38) and a metasurface layer. To be consistent with the analytical studies, the dielectric is assumed to be lossless and the conductive material is selected to be perfect electric conductor (PEC), such that the only loss mechanism for the simulated metasurface antenna is radiation. Excitation of the metasurface antenna is implemented at the first port (left) of the antenna while the second port (right) of the metasurface is terminated in 50 $\Omega$ to absorb the remaining guided-mode reference and prevent it from being reflected back to the waveguide. The metamaterial elements etched onto the metasurface layer are sub-wavelength sized, complementary slot-shaped elements that couple to the magnetic field of the launched quasi-TEM mode at the input port \cite{PhysRevApplied.8.054048}. The length and width of the metamaterial elements are $\lambda_g$/4 and $\lambda_g$/30, respectively, where $\lambda_g$ denotes the wavelength inside the dielectric at the center frequency, 10 GHz. 

The distribution of the metamaterial elements is selected in such a way that the sampling of the guided-mode reference is achieved at the phase-matched points across the aperture for a given steering angle while the slot-shaped metamaterial elements exhibit a uniform coupling strength at 10 GHz. Therefore, the simulated design closely resembles the proposed Euclidean modulation ($y_0=-0.5$) in terms of the phase and amplitude characteristics of the resultant aperture wave-front, providing a performance comparison in terms of the far-field radiation pattern characteristics. Whereas the Euclidean modulation can achieve the desired phase and amplitude profile by sampling the aperture at arbitrary selected points across the aperture, the adopted technique for the full-wave simulations realizes the desired phase and amplitude profile at phased-matched points across the aperture. Using slot-shaped weakly-resonant metamaterial elements brings the advantage of reducing the excessive losses caused by highly resonant elements, increasing the radiation efficiency of the antenna. 

\begin{figure}[h!]
\centering
\includegraphics[width=7cm,trim=1cm 0cm 1cm 0cm]{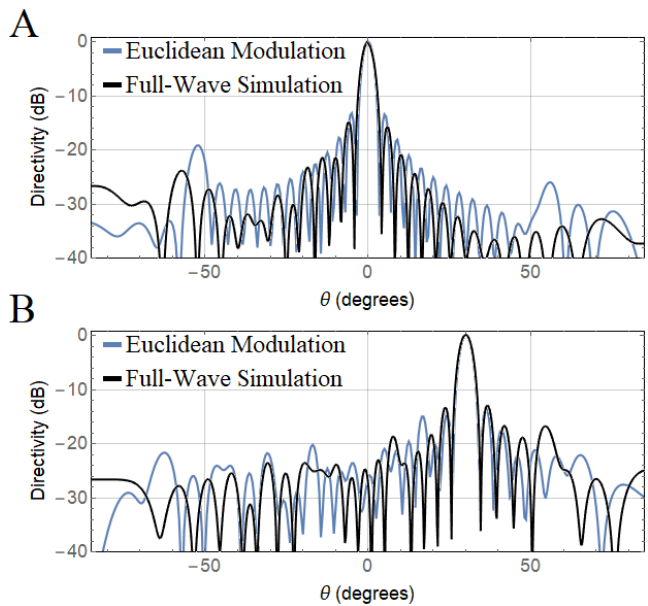}
\caption{Analytical (Euclidean modulation) and simulated directivity patterns (a) broadside radiation (b) steered to 30 degrees.}
\label{fig:Analytical_Numerical_Pattern_Comparison}
\end{figure}

The simulated radiation patterns of the metasurface antenna are shown in Fig. \ref{fig:Analytical_Numerical_Pattern_Comparison}, along with the calculated analytical radiation patterns for Euclidean modulation. Analyzing the simulated radiation patterns of the metasurface antenna, the sidelobe levels are measured to be -14.9 dB for broadside radiation ($\theta$=0 deg.) and -13.1 dB for steered radiation ($\theta$=30 deg.) at 10 GHz. Comparing the radiation patterns in Fig. \ref{fig:Analytical_Numerical_Pattern_Comparison}, the agreement in the pattern envelopes between the analytically calculated Euclidean modulation directivity patterns and full-wave directivity patterns is evident. Another important figure of merit for the simulated metasurface antennas is the antenna radiation efficiency, which is measured to be above 80\% for all of the simulated designs.     

\section{Conclusion}
\label{sec:conclusion}
We have presented a family of complex plane mappings that allow the polarizability distribution of a set of  Lorentzian-constrained metamaterial elements to be determined, parametrized by a focal point $y_0$ that lies along the negative imaginary axis. Certain values of this parameter lead to highly optimized solutions for the effective polarizabilities of the metamaterial elements that can be readily applied in many different contexts; this is a nontrivial problem due to the fact that metasurface elements only grant control over the resonant frequency, which results in coupled shifting of the magnitude and phase of each radiator. In particular, Euclidean modulation ($y_0=-0.5$) is particularly easy to implement and can be applied more generally to a wide class of metasurfaces.

\section{Acknowledgments}
This work was supported by the Air Force Office of Scientific Research (AFOSR, Grant No. FA9550-18-1-0187).

\appendix*
\section{Tensor form of the polarizability}

If it is desired to extend the above analysis to metasurfaces that can produce arbitrarily polarized fields, a collection of metamaterial elements with magnetic dipole moments directed arbitrarily within the aperture plane must be considered. In this case, we assume the wave propagates in any direction away from the aperture and that, as described above, the required surface current is as described by Eq. \ref{eq:FundamentalDesignEqn1d}. Then, the magnetic surface current can be written as a magnetic surface susceptibility tensor \(\bar{\chi}_M\) multiplied by the magnetic field of the feed wave \(\textbf{H}_f\):
\begin{equation}\label{eq:ProducingSurfaceCurrent}
\textbf{K}_M(\textbf{r})=-j\omega \mu_0 \bar{\chi}_M(\textbf{r})\textbf{H}_f(\textbf{r}).
\end{equation}

To allow the antenna to produce a beam with arbitrary polarization, both eigenvectors of the surface susceptibility tensor need to be orthogonal and the susceptibility controlled along the two different directions. The effective surface susceptibility tensor can be implemented by imposing two lattices of metamaterial scattering elements, where each element in each lattice can be excited with a magnetic dipole moment in only one direction, and the directions of the dipoles in the two lattices are orthogonal \cite{karamanos2012polarizability}.  

To derive how these lattices of magnetic dipoles must be controlled, consider two lattices where the positions of the dipoles in the first lattice are \(\textbf{r}_i\) with orientations \(\hat{\nu}_i\) and where the positions of the dipoles in the second lattice are also \(\textbf{r}_i\), but with orientations \(\hat{\mu}_i\) such that \(\hat{\nu}_i\cdot\hat{\mu}_i=0\).  Then the surface susceptibility tensor can be generally written as 
\begin{equation}\label{eq:SusceptibilityTensor}
\bar{\chi}_M(\textbf{r}_i)=\chi_1(\textbf{r}_i)\hat{{\nu}}_i\otimes\hat{{\nu}}_i+\chi_2(\textbf{r}_i)\hat{{\mu}}_i\otimes\hat{{\mu}}_i
\end{equation}
\noindent where the symbol \(\otimes\) is used to designate the tensor product. 

Under the assumptions of weak coupling assumed above, the surface susceptibility at position \(\textbf{r}_i\) is related to the magnetic polarizability \(\alpha_i^1\) of the element in the first lattice at position \(\textbf{r}_i\) by \(\alpha^1_i=\Lambda^2\chi_1(\textbf{r}_i)\), while the polarizability of the element in the second lattice is likewise \(\alpha^2_i=\Lambda^2\chi_2(\textbf{r}_i)\). Here, $\Lambda$ is the spacing between metamaterial elements. Using this relationship, together with Eqs. \ref{eq:FundamentalDesignEqn1d}, \ref{eq:SusceptibilityTensor} and \ref{eq:ProducingSurfaceCurrent}, the required polarizabilities of the dipoles are
\begin{subequations}
\begin{align}
\alpha_1^i&=\frac{j\Lambda^2}{Z_0 k}\textbf{E}_0\cdot(\hat{n}\times\hat{\nu}_i)\left(\frac{e^{-j\textbf{k}_b\cdot\textbf{r}_i}}{\textbf{H}_f\cdot\hat{\nu}_i}\right) \\
\alpha_2^i&=\frac{j\Lambda^2}{Z_0 k}\textbf{E}_0\cdot(\hat{n}\times\hat{\mu}_i)\left(\frac{e^{-j\textbf{k}_b\cdot\textbf{r}_i}}{\textbf{H}_f\cdot\hat{\mu}_i}\right).
\end{align}
\end{subequations}

\noindent Here we have used the dispersion relation $\omega=c k$, where $c^{-2}=\epsilon_0 \mu_0$ and $Z_0^{2}=\mu_0/\epsilon_0$.

Choosing the specific form \(\textbf{H}_f=\textbf{h}_fe^{-j\beta y}\), where \(\beta\) is the propagation constant and the waveguide is assumed to lie in the \(y\)-direction, the requirement on the polarizabilities to produce an ideal beam pattern is
\begin{subequations}
\begin{align}
\alpha_1^i&=\frac{j\Lambda^2}{Z_0 k}\left(\frac{\textbf{E}_0\cdot(\hat{n}\times\hat{\nu}_i)}{\textbf{h}_f\cdot\hat{\nu}_i}\right)e^{-j(\textbf{k}_b-\beta\hat{\textbf{y}})\cdot\textbf{r}_i} \label{eq:LinearWgIdealModulation1} \\
\alpha_2^i&=\frac{j\Lambda^2}{Z_0 k}\left(\frac{\textbf{E}_0\cdot(\hat{n}\times\hat{\mu}_i)}{\textbf{h}_f\cdot\hat{\mu}_i}\right)e^{-j(\textbf{k}_b-\beta\hat{\textbf{y}})\cdot\textbf{r}_i}. \label{eq:LinearWgIdealModulation2}
\end{align}
\end{subequations}

It should be noted that the techniques described above can easily be adapted to antennas in which the guided wave is restricted to propagate in two directions, or two-dimensional (2D) waveguide-fed metasurfaces. For example, for a 2D metasurface antenna, if the feed structure is a planar waveguide that is centrally fed, the feed wave follows the form \(\textbf{H}_f=h_f\hat{\theta}H^{(2)}(\beta r)\approx h_f\hat{\theta}e^{-j\beta r}/\sqrt{\beta r}\).
\begin{subequations}
\begin{align}
\alpha_1^i&=\frac{j\Lambda^2}{Z_0 k}\left(\frac{\textbf{E}_0\cdot(\hat{n}\times\hat{\nu}_i)}{{h}_f\hat{\theta}\cdot\hat{\nu}_i}\right)\sqrt{\beta r_i}e^{j({k}_0-\beta){r}_i} \label{eq:PlanarWgIdealModulation1} \\
\alpha_2^i&=\frac{j\Lambda^2}{Z_0 k}\left(\frac{\textbf{E}_0\cdot(\hat{n}\times\hat{\mu}_i)}{{h}_f\hat{\theta}\cdot\hat{\mu}_i}\right)\sqrt{\beta r_i}e^{j({k}_0-\beta){r}_i}. \label{eq:PlanarWgIdealModulation2}
\end{align}
\end{subequations}
In this case the required polarizability rotates through all possible phases, but increases as \(\sqrt{r}\) to compensate for the natural spread in the feed wave through the aperture. As in the 1D case, the requirement for an ideally radiating antenna is that all possible phases of the polarizability of the scattering elements be available. For the Lorentzian-constrained metamaterial elements, Euclidean modulation provides useful solutions for beam forming with a 2D aperture.

\bibliography{LorentzianConstrainedPol}

\end{document}